# Energy gap evolution in the tunneling spectra of $Bi_2Sr_2CaCu_2O_{8+\delta}$

Renier M. DIPASUPIL, Migaku ODA, Naoki MOMONO and Masayuki IDO

*Department of Physics, Hokkaido University, Sapporo 060-0810, Japan*

On the basis of the tunneling spectra in Bi2212/vaccum/Bi2212 junctions fabricated using STM, we report that, in the electronic excitations, there exist two kinds of pseudogaps; one with a characteristic energy comparable to the superconducting (SC) gap and another one that is 3 to 4 times larger. The smaller energy-scale pseudogap (SPG) develops progressively below temperature $T^*$, which nearly scales with the SC gap amplitude $\Delta_0$ at $T \ll T_c$, in addition to the larger energy-scale pseudogap (LPG), which already exists above $T^*$. The SPG smoothly develops into the SC-state gap with no tendency to close at $T_c$.

## 1. Introduction

Since the discovery of high-$T_c$ cuprates, energy gap $\Delta(\boldsymbol{k})$ in the superconducting (SC) state (the SC-state gap) has been studied extensively to elucidate the SC mechanism. For the symmetry of $\Delta(\boldsymbol{k})$, most experimental results are consistent with a $d$-wave gap, which has lines of nodes at the directions 45° from the Cu-O bonds or the $k_x$ and $k_y$ axes in the $\boldsymbol{k}$-space ($\Delta(\boldsymbol{k})=\Delta_0(\cos k_x - \cos k_y)$). For the amplitude $\Delta_0$ of $\Delta(\boldsymbol{k})$ at $T \ll T_c$, it was demonstrated in tunneling and photoemission experiments on $Bi_2Sr_2CaCu_2O_{8+\delta}$ (Bi2212) and $La_{2-x}Sr_xCuO_4$ (La214) that $\Delta_0$ increases monotonically with the lowering of the hole-doping level $p$, although $T_c$ decreases after exhibiting a maximum at a certain doping level, $p_o$; that is, $T_c$ does not scale with $\Delta_0$.[1-7] Interestingly, in the underdoped (UD) region ($p<p_o$), $T_c$ nearly scales with the product of $\Delta_0$ and $p$, $p\Delta_0$; the characteristic energy in determining $T_c$ or the effective SC gap is $\sim p\Delta_0$, instead of $\Delta_0$ in conventional BCS superconductors, suggesting that the SC transition mechanism will be seriously modified from the BCS type.[8-10] As is well known in UD high-$T_c$ cuprates, the normal-state electronic property is characterized by a significant suppression of the spectrum at low energies, that is, the so-called "pseudogap." This tempts us to suppose that the development of the pseudogap will strongly affect the SC transition mechanism. Therefore, the pseudogap has also been studied extensively by spectroscopic techniques, in particular for Bi2212, where very clear crystal surfaces, indispensable for such experiments, can be obtained by cleaving. However, no consensus has been achieved for some interesting properties of the pseudogap, as described below.

According to angle-resolved photoemission spectroscopy (ARPES) for Bi2212, the SC-state gap exhibits no tendency to close at $T_c$ although it is slightly reduced with the increase of $T$, and smoothly connects with a pseudogap above $T_c$, the so-called "small (or strong) pseudogap (SPG)", which is gradually suppressed with the increase of $T$ and disappears around temperature $T^*$.[11,12] On the other hand, it was reported in scanning tunneling spectroscopy (STS) experiments on Bi2212 and neutron inelastic scattering experiments on $YBa_2Cu_3O_{6+\delta}$ that, especially at low doping levels, the SPG remains, at least up to the highest temperature examined ($T \sim 300$ K), much higher than $T^*$ in the ARPES experiments.[5,11-14] More contradictory results were recently reported in interlayer tunneling experiments on Bi2212; it was claimed that the SC-state gap, which is rapidly reduced as a function of increasing $T$, tends to close at $T_c$, and a $T$-independent pseudogap, which has almost the same energy scale as the SC-state gap at $T \ll T_c$ but no connection with the superconductivity, exists in the normal state.[15,16] Furthermore, it was suggested by recent angle-integrated photoemission spectroscopy (AIPES) that another pseudogap with a larger characteristic energy, the so-called "large (or weak) pseudogap (LPG)", would also exist over a wide $T$ range including high temperatures above $T^*$;[17] however, no clear evidence for the existence of the LPG at high temperatures was provided in STS experiments.[5,13]

In this study, it is clearly demonstrated that in the normal-state tunneling spectra of Bi2212 there exist two kinds of pseudogaps, SPG and LPG, whose characteristic energies are comparable to and 3 to 4 times larger than the SC-state gap, respectively. We report that the SPG develops progressively below the mean-field characteristic temperature $T_{co}$ ($=2\Delta_0/4.3k_B$) for $d$-wave superconductors, in addition to the LPG, which already exists above $T_{co}$, and smoothly develops into the SC-state gap below $T_c$. Furthermore, we report that the high-energy quasiparticle excitations related to the LPG will also be modified across $T_c$.

## 2. Experiments

Bi2212 single crystals were grown by the TSFZ method. The SC critical temperature $T_c$ of the as-grown crystals was ~90 K, indicating that their hole-doping level was nearly optimal. To obtain underdoped (UD) and overdoped (OD) samples, the as-grown crystals were annealed at 700 in an atmospheric oxygen-nitrogen mixed gas with a low oxygen concentration of 0.1% and in a high-pressure (20 atm) oxygen gas, respectively. The doping level $p$ was basically estimated from the $T_c$ value so that it fell on the $T_c$-$p$ curve determined by Gröen *et al*.[18] Since the $T_c$-$p$ curve is parabolic, a $T_c$ value smaller than its maximum gives two doping levels, which are under- and overdoped, respectively. Whether it is under- or overdoped is con-firmed from the *T*-dependence of the magnetic susceptibility, which changes systematically as a function of *p*.

In the present tunneling experiments, Bi2212/vaccum/Bi2212 (BVB) junctions, as illustrated in Fig. 1 (a), were fabricated in a scanning tunneling microscope (STM) system using the junction fabrication technique that was first achieved in a point-contact tunneling spectroscopy system by Miyakawa *et al.*[4] The STM tip contacted a Bi2212 crystal and was then spaced slightly away from the crystal; a small part of the crystal, whose size was on the order of 10 μm, was cleaved, sticking to the tip, thus providing a fresh BVB junction, which is of the superconductor/insulator/superconductor (SIS) type below $T_c$. This junction fabrication was performed in low-pressure ($10^{-2} \sim 10^{-3}$ torr) helium gas at a low temperature (~10 K) to avoid oxygen deficiency in the cleaved surfaces composing the BVB junction.

## 3. Results and discussion

Figures 1 (b) and (c) show a typical example of low-temperature ($T \ll T_c$) current-voltage (*I-V*) characteristics in BVB junctions and the corresponding d*I*/d*V*-*V* curve (tunneling spectrum), respectively. In the present experiments, the values of junction conductance (or resistance) at high voltages were $10^{-4} \sim 10^{-5}$ $\Omega^{-1}$ (or 10 ~ 100 kΩ). In the inset in Fig. 1 (b), the *I-V* curve is magnified in the *V* range around the zero-bias voltage. One can see in the inset that the *I-V* curve exhibits a small current step at *V*~0, which is due to the Josephson coupling between the SC Bi2212 crystals in the BVB junction. The Josephson current corresponds to the sharp peak at *V*~0 in the tunneling spectrum (Fig. 1 (c)).

In SIS-type junctions such as BVB, the tunneling spectrum d*I*/d*V* is given by the following equation at low temperatures ($T \ll T_c$):

$$dI/dV \propto \int_0^{eV} N_s(E) N_s(E - eV) dE,$$

where $N_s(E)$ is the quasiparticle density of states (spectrum) and the Fermi energy $E_F$ is taken to be zero. The SIS-type tunneling spectrum is not the quasiparticle spectrum itself, which can be directly measured by superconductor/insulator/normal-metal (SIN) type tunneling spectroscopy. However, the positions of the coherence peaks in the SIS-type tunneling spectrum, arising from the divergence of $N_s(E)$ at the gap edges $E = \pm \Delta_0$, are little affected by the thermal broadening of $N_s(E)$ except for the vicinity of $T_c$; they appear at voltages $V \sim \pm 2\Delta_0/e$ over a wide *T* range ($T \lesssim T_c$), as is well known (Fig. 1 (c)).[4] Furthermore, BVB junctions are rather stable for temperature variation, in addition to the advantage that they consist of freshly cleaved surfaces, which are expected to have no oxygen deficiency. These are the reasons why we used BVB junctions to examine the *T* dependence of the Bi2212 tunneling spectrum.

In Fig. 2, tunneling spectra, which were measured in the BVB junctions of OD (*p*~0.21) and UD (*p*~0.14) samples with the same $T_c$ (82 K), are shown for selected temperatures. The conductance d*I*/d*V* (σ) data at each temperature were obtained while keeping the temperature constant in the course of heating. In the course of heating, the interval between the two crystals in the BVB junction or the junction barrier thickness changed gradually because of thermal distortions of the STM sample unit, and was therefore controlled in the present study so that the d*I*/d*V* value at a voltage in the range from –150 mV to –200 mV was kept nearly constant within ±10% of the average over the entire *T* range examined. To remove even the small extrinsic fluctuation of the data as a function of *T*, the spectrum at each temperature was normalized with its value at the voltage where the junction barrier thickness was controlled.

In the lowest-temperature (7 K) spectrum of the OD sample (Fig. 2 (a)), one can see a very clear SC gap structure; it exhibits sharp coherence peaks at voltages $|V| = V_p$ (=50 mV), and the gap magnitude $2\Delta_0$ at $T \ll T_c$ is estimated to be 50 meV from the peak voltages. In addition, the spectrum also exhibits a clear hump and dip structure outside the energy gap, which will be discussed in the latter part of this paper, together with a small peak at *V*~0. The zero-bias peak is due to a small step of the Josephson current, as mentioned above. The energy gap in the SC state, the SC-state gap, is slightly reduced with the increase of *T* but exhibits no tendency to close at $T_c$, and even in the normal state, a gap-like suppression of the spectrum still remains in an energy range comparable to the SC-state gap, although the steep enhancement of the spectrum around $|V| = V_p$ or its coherence feature disappears. Similar behavior was also reported in previous tunneling experiments on OD Bi2212 samples.[19,20] Here, the gap-like structure with almost the same energy scale as the SC-state gap is referred to as a "small (or strong) pseudogap (SPG)".

Figure 3 (a) is a three-dimensional (3D) illustration of the OD spectrum in the *T* range from 60 K to 180 K, which is generated from the d*I*/d*V* data in Fig. 2 (a) to provide an easy understanding of the energy gap evolution. In the bottom panel, furthermore, the spectrum is compressed along the vertical axis to see its entire change over a wide *T* range. At high temperatures, the spectral shape is characterized by a broad bump with its maximum at around *V*=0, and pseudogap-like behavior cannot be seen clearly. As the temperature is lowered, the SPG develops progressively below ~120 K in the bump-shaped spectrum, and then evolves smoothly into the SC-state gap at $T_c$.

On the other hand, Fig. 3 (b) is a 3D illustration of the UD spectrum generated from the d*I*/d*V* data in Fig 2 (b). At high temperatures, the UD spectrum is qualitatively different from the OD spectrum; it exhibits a very broad hump or saturation behavior at around $|V| \sim 200$ mV and is gradually suppressed toward *V*=0 for $|V| \lesssim 200$ mV (Figs. 2 (b) and 3 (b)). Such a gradual suppression of the spectrum in a large energy region, whose size is 3 to 4 times larger than the SC-state gap, is referred to as a "large (or weak) pseudogap (LPG)." One can also see in the 3D illustration of the UD spectrum (Fig. 3 (b)) that an SPG, which is defined as a sharper suppression of the spectrum in the energy range comparable to the SC-state gap, develops progressively below ~180 K, in addition to the LPG, which already exists at higher temperatures, and evolves into the SC-state gap below $T_c$. A similar result on the temperature evolution of the energy gap was obtained for a more UD sample with $p\sim0.12$ and $T_c=77$ K, as shown in Fig. 3 (c) (3D illustration of the tunneling spectrum). Furthermore, such an energy gap evolution can be seen in the STS spectra of a UD Bi2212 sample with $T_c=83$ K, reported by Renner *et al.*, as well.[5]

Figures 4 (a), (b) and (c) show the $T$ dependences of the zero-bias conductance $\sigma(0)$, reflecting the density of states at the Fermi level, in the tunneling spectra for three different doping levels, $p\sim0.12$, $\sim0.14$ and $\sim0.21$ (Figs. 3 (a), (b) and (c)), respectively. The arrows in the figures represent the $T_c$ values, determined from dc-SQUID diamagnetic susceptibility measurements. For $p\sim0.14$, the $\sigma(0)$ data measured in another BVB junction above $\sim100$ K were also plotted, together with those in Fig. 3 (b), to demonstrate that the $T$ dependences of $\sigma(0)$ in both junctions are qualitatively consistent with each other. At high temperatures, $\sigma(0)$ tends to decrease gradually as a function of lowering $T$ in the UD samples, while it is nearly constant in the OD sample. Such a gradual decrease of $\sigma(0)$ in the UD samples at high temperatures will be due to the gradual development of the LPG. For all the doping levels, $\sigma(0)$ starts to decrease more sharply with crossover-like behavior at around temperature $T^*$, below which the SPG develops progressively, and drops off below $T_c$. The $T^*$ value, defined as in Fig. 4, is $\sim200$ K, $\sim180$ K and $\sim120$ K for $p\sim0.12$, $\sim0.14$ and $\sim0.21$, respectively; it becomes lower at higher doping levels. Furthermore, the degree of development in the SPG, defined as the difference ($\Delta\sigma$) of $\sigma(0)$ between $T^*$ and $T_c$, becomes smaller at higher doping levels, in accordance with the decrease of $T^*$.

In Fig. 5, the $T^*$ data in the tunneling experiments are plotted as a function of $p$, together with those in ARPES experiments and the data on the spin-gap temperature, around which a pseudogap in the spin excitations, the so-called "spin gap", has been demonstrated in NMR experiments to open up with a crossover-like development.[12,21,22] The low-temperature ($T \ll T_c$) energy gap $2\Delta_0$ data obtained in the present study, whose $p$ dependence is consistent with the results of previous tunneling experiments,[2-4] are shown in Fig. 5, as well. In the inset, furthermore, the reduced gap $2\Delta_0/k_BT^*$, is plotted as a function of $p$. One can see in this figure that the tunneling $T^*$, which is comparable to the ARPES and NMR $T^*$, nearly scales with $2\Delta_0$ over the entire $p$ range examined (see the inset).

It is well known from the BCS theory for $d$-wave superconductors that $2\Delta_0=4.3k_BT_{co}$, where $T_{co}$ is the mean-field critical temperature.[23,24] A similar relation holds for the onset temperature of spinon pairing in the mean-field slave-boson $t$-$J$ model, where superconductivity occurs as holons or holon pairs condense in addition to the formation of spinon pairs.[25,26] Thus, in Fig. 5, we plot temperature $T_{co}$ ($=2\Delta_0/4.3k_B$) obtained from the tunneling gap. As shown in the figure, SPG temperature $T^*$ is comparable to $T_{co}$; in other words, $2\Delta_0/k_BT^*$ is close to the mean field value. Very recently, Kugler *et al.* have also demonstrated that relation $T^*\sim T_{co}$ holds for an OD $Bi_2Sr_2CuO_{6+\delta}$ sample with $T_c=10$ K, $T^*\sim68$ K and $2\Delta_0=24$ meV.[27] Furthermore, it has been found in La214 that the electronic specific heat, reflecting the density of states at the Fermi level, exhibit pseudogap-like behavior below $\sim T_{co}$, as well.[3,28] It is therefore considered that in high-$T_c$ cuprates, the SPG develops progressively below $\sim T_{co}$. This, together with the fact that the SC-state gap seems to grow from the SPG, suggests that the SPG will be some kind of precursor of superconductivity.[25,26,29-33] Interestingly, the in-plane electric resistivity and uniform magnetic susceptibility start to decrease slightly more sharply below $\sim T_{co}$, which is consistent with the contention that some kind of precursor of superconductivity will occur below $T^*\sim T_{co}$.[2,3]

For the electronic phase diagram, as in Fig. 5, it should be noted that in the $T$ region above $T^*$, there exists another crossover temperature $T_{max}$, around which the magnetic susceptibility $\chi$ starts to decrease gradually with the lowering of $T$ after exhibiting a broad peak.[2] Recent AIPES experiments reported that the LPG, which remains open even at high temperatures above $T^*$ ($\sim T_{co}$), develops progressively below $\sim T_{max}$.[17] It has been considered on the basis of analyses of the $\chi$-$T$ curve that $k_BT_{max}$ is a characteristic energy for the effective antiferromagnetic (AF) interaction of Cu-spins and that the decrease in $\chi$ arises from the gradual development of AF spin fluctuation.[2,34,35] Temperature $T_{max}$ is rapidly reduced with the increase of $p$, and for the OD $p\sim0.21$ sample, the magnitude of the $\chi$-decrease below $T_{max}$ ($\sim200$ K), giving a measure of the degree of development in the AF spin fluctuation, is quite small even at around 100 K just above $T_c$. On the assumption that the LPG will be related to the AF spin fluctuation, it is therefore expected that, for the OD sample, the degree of development in the LPG will be quite small in the normal state as well. This is qualitatively consistent with the present result that in the OD spectrum (Figs. 2 (a) and 3 (a)), LPG-like behavior is unclear, at least, in the normal state, although a hump structure, implying the existence of an LPG,[7] is clearly seen at low temperatures well below $T_{max}$.

In the UD spectrum for $p\sim0.14$ (Fig. 3 (b)), one can see that across $T_c$, some changes occur in the high-energy region outside the SC-state gap, in addition to the rapid growth of the SC-state gap from the SPG at lower energies. Below $\sim T_c$, a dip structure, which has been considered to be due to the strong couplings of quasiparticles to magnon or phonon collective mode,[36-39] grows gradually just outside the SC-state gap, as clearly shown in Fig. 6, and the spectral weights at high energies seem to move towards low energies; thus, the hump structure becomes much more visible, accompanied by a small shift of its position toward lower energies. Furthermore, the dip and hump positions shift slightly towards higher energies at low temperatures, which seems to correlate with the increase of the coherence peak energy or the SC-state gap magnitude $2\Delta_0$. The clear peak-dip-hump structure has been observed in another tunneling experiments and demonstrated to be a characteristic feature of the ARPES spectra around ($\pm\pi$, 0) and (0, $\pm\pi$) in the SC state, as well.[4-7,11,12] On the other hand, the change in the high-energy hump structure across $T_c$ has not been confirmed in ARPES experiments. Therefore, it could be explained from the experimental point of view, for example, as follows: an extrinsic background, which made the clear hump structure indistinct, is superposed in the tunneling spectra above $T_c$. However, such a change in the UD tunneling spectrum across $T_c$ is reproducible in our BVB junction experiments (see Fig. 3 (c) for another UD sample) and can also be seen in the Bi2212 tunneling spectra measured in the same type of junctions by Miyakawa *et al.*[4,7] and in recent point-contact junctions by Oki *et al.*[40] Another explanation for the change in the high-energy hump structure across $T_c$ could also be made on the basis of a possible intrinsic origin, for example, in terms of the AF spin fluctuations; in the SC state below $T_c$, where a coherent spin-singlet state is realized, the AF spin fluctuations will be suppressed, leading to the observed change in the high-energy spectrum responsible for the LPG.

**4. Summary**

In summary, the tunneling spectra of Bi2212 were measured over a wide $T$ range in BVB junctions fabricated using STM, and it was clearly demonstrated that, in the electronic excitations, there exist two kinds of pseudogaps, SPG and LPG; the former one (SPG) with a characteristic energy comparable to the SC-state gap and the latter one (LPG) that is 3 to 4 times larger. The SPG, which develops progressively below the mean-field characteristic temperature $T_{co}$ for $d$-wave superconductors and then evolves into the SC-state gap at $T_c$, will be some kind of precursor of superconductivity. On the other hand, the LPG, which remains open even at high temperatures above $T_{co}$, seems to be related to the crossover behavior of magnetic susceptibility around $T_{max}$ ($>T_{co}$), which arises from the gradual development of AF spin fluctuations. It was also found in UD Bi2212 that, in accordance with the SC transition, the characteristic feature of the spectrum at high energies changes from a broad hump structure to a clear hump and dip structure accompanied by a shift of the hump position, in addition to the rapid growth of the SC-state gap from the SPG at low energies. Further studies on such a modification in the quasiparticle excitations across $T_c$ will be needed to understand the SC transition mechanism and/or the pairing mechanism more profoundly.

**Acknowledgments**

The authors would like to thank Prof. J. F. Zasadzinski and Dr. K. E. Gray for their helpful information about BVB junction fabrication and useful discussions. They are also grateful to Prof. F. J. Ohkawa and Dr. A. V. Balatsky for valuable discussions. This work was supported in part by Grant-in-Aids for Scientific Research on Priority Area (No. 407, "Novel Quantum Phenomena in Transition Metal Oxides") and on Projects (Nos. 12640327, 12740191 and 13440104) from the Ministry of Education, Culture, Sports, Science and Technology of Japan, and by the Kurata Foundation.

**Correspondence**: M. Oda (moda@sci.hokudai.ac.jp)

## Figure Captions

Fig. 1: Schematic diagram of BVB junction (a), *I-V* characteristic at 5 K in a UD BVB junction (b) and the corresponding tunneling spectrum (c).

Fig. 2: Tunneling spectra (d*I*/d*V*-*V* curves) in the BVB junctions of OD $p\sim 0.21$ and UD $p\sim 0.14$ Bi2212 samples. The spectrum at each temperature is normalized with the value at −150 mV. The zero level in the vertical axis is only for the lowest-temperature spectrum, and the other spectra are shifted upwards for clarity.

Fig. 3: 3D illustrations of the tunneling spectra for $p\sim 0.21$, 0.14 and 0.12, which were generated by using 3D imaging software. The $T_c$ value, indicated in this figure, was determined from dc-SQUID diamagnetic susceptibility measurements.

Fig. 4: *T* dependence of normalized $\sigma(0)$ (zero-bias conductance). The $\sigma(0)$ data for $p\sim 0.12$, 0.14 and 0.21, represented by closed circles, are plotted from the tunneling spectra in Figs. 3 (a), (b) and (c), respectively. For $p\sim 0.14$, the $\sigma(0)$ data measured in another BVB junction above ~100 K are also shown by open circles, together with those in Fig. 3 (b), to demonstrate that the *T* dependences of $\sigma(0)$ in both junctions are qualitatively consistent with each other. The arrows represent the $T_c$ values determined from dc-SQUID diamagnetic susceptibility measurements.

Fig. 5: Dependences on $2\Delta_0$, $T^*$, $T_{co}$ and $T_c$ on $p$ in Bi2212. Temperature scale is taken so that the $2\Delta_0$ data give the $T_{co}$ values. The inset shows the $p$-dependence of $2\Delta_0/k_BT^*$, which was obtained from the tuneling $T^*$ and low-temperature ($T<<T_c$) tunneling gap $2\Delta_0$.

Fig. 6: Temperature dependence of the dip depth defined as in the inset. The data were obtained from two UD ($p\sim 0.14$) BVB junctions and normalized with the lowest-temperature value.

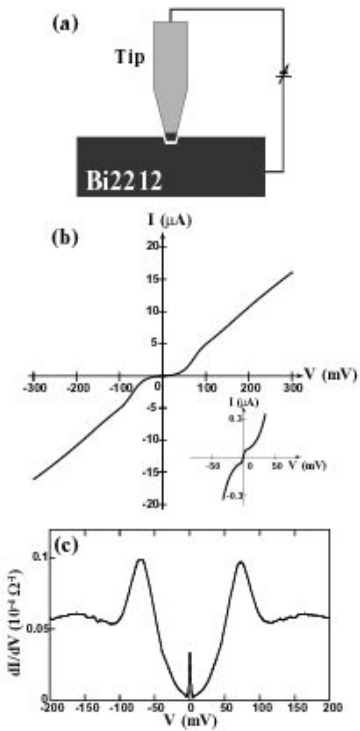

Fig. 1

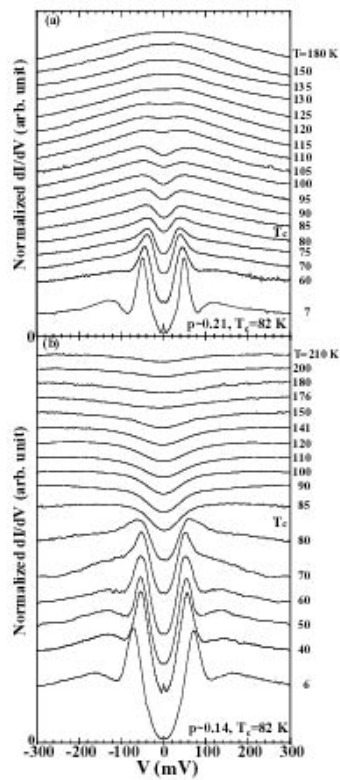

Fig. 2

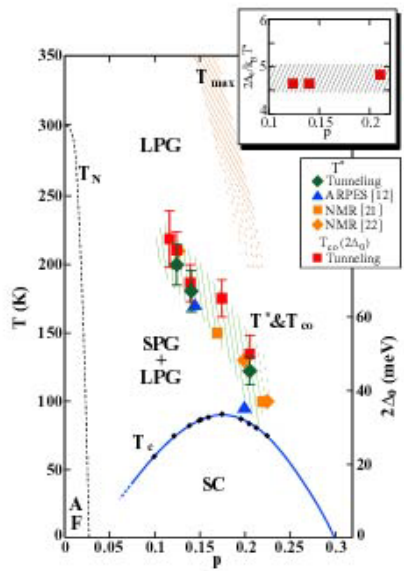

Fig. 3

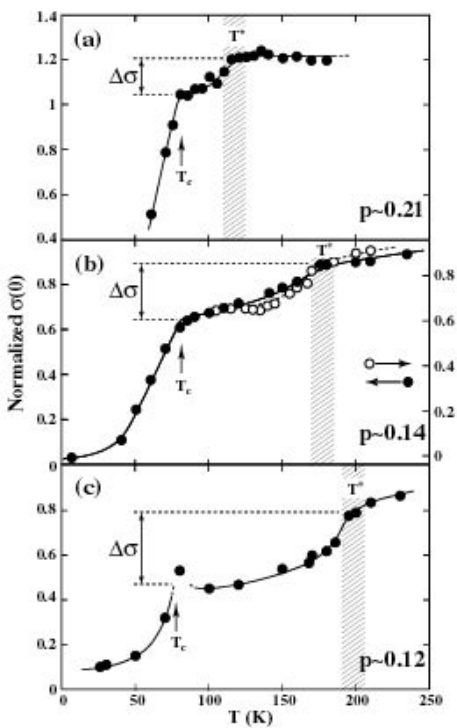

Fig. 4

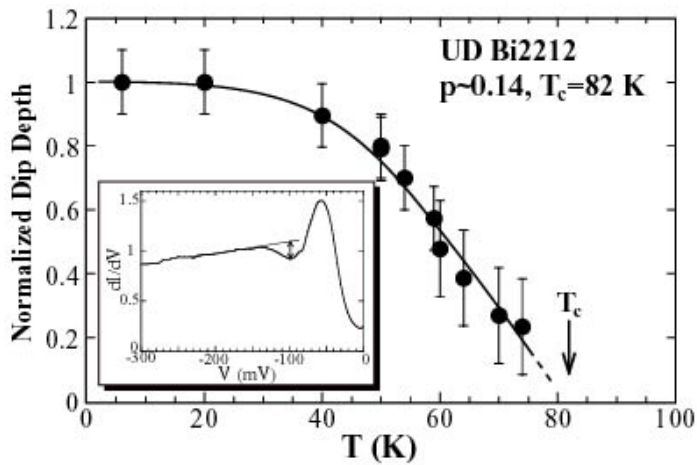

Fig. 6

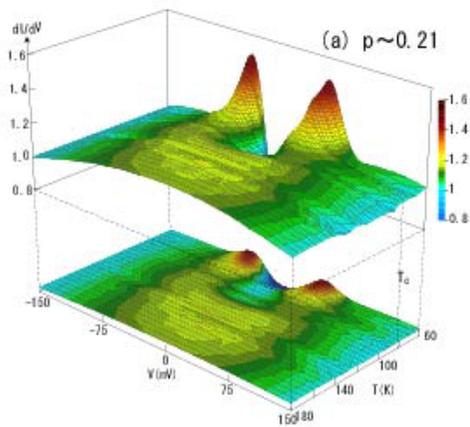
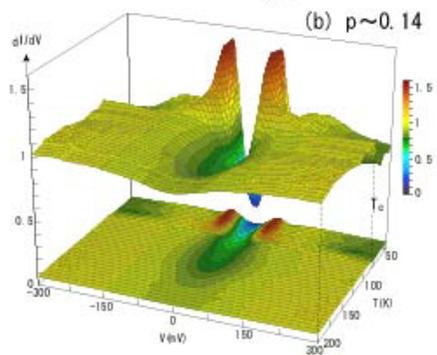
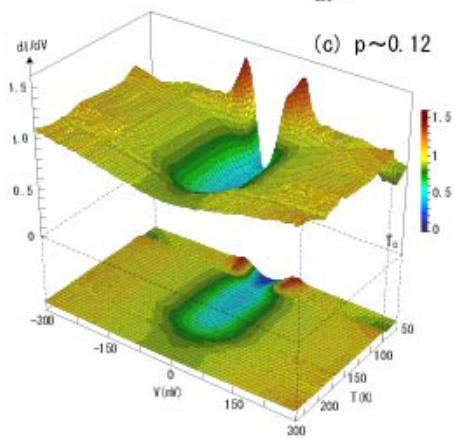

Fig. 3